\documentclass[]{mn2e}
\usepackage{graphics}

\title[3.4-$\mu$m band in absorption and emission]{On the carriers of the 3.4-$\mu$m absorption and emission bands, and their evolution}
\author[R. Papoular]{Renaud Papoular$^{1}$\thanks{E-mail:papoular@wanadoo.fr}\\
$^{1}$Service d'Astrophysique and Service de Chimie Moleculaire,
CEA Saclay, 91191 Gif-s-Yvette, France}
\begin{document}

\date{Accepted . Received ; in original form }

\pagerange{\pageref{firstpage}--\pageref{lastpage}} \pubyear{2002}

   \maketitle
\label{firstpage}

\begin{abstract}
The 3.4-$\mu$m feature is one of the celebrated Unidentified InfraRed bands. However, it may be singled out for coming with a great variety of intensities relative to the 3.3-$\mu$m feature, depending on the environment of its emitters: it is nearly invisible in emission spectra from the Diffuse Interstellar Medium, and it is overwhelming in Young Stellar Objects.

In this work, based on the results of chemical analysis and simulation of kerogens and immature coals, a large number of chemical structures carrying the 3.4-$\mu$m feature were studied by means of computer simulation codes. Further selection criteria were the integrated strength of the absorption lines in the aliphatic stretchings wavelength band, weak IR activity in the aromatic stretching band and absence of notable activity outside the astronomical UIBs (Unidentified Infrared Bands). Most of the structures that were retained can be classed as branched, short and oxygen-bridged CH$_{2}$ chains, and naphtenic chains. Combinations of their absorption spectra deliver spectra comparable to those observed in the sky. 

Absorption spectra were derived from Normal Mode Analysis. Emission spectra of the same structures were computed by monitoring their overall dipole moment as they vibrate freely in vacuum after excitation. These spectra were then combined in suitable proportions, together with those of aromatic structures, so as to simulate various typical near IR emission spectra observed in the sky. 

Examples of graphical fits to different types of absorption and emission spectra are displayed.  This study also underscores the facts that: a) no band can be ascribed to a single fingerprint line, b) the activity of a given functional group contributing to a band depends on its molecular environment and, therefore, cannot be predicted on the sole basis of the corresponding integrated line intensity.

A review of the spatial distribution, in our and external galaxies, of the various classes of CH stretching spectra in the light of these results suggests the following scenario for the evolution of the feature carriers: short chains are preferentially formed in the winds of AGB stars and novae, but also in the interstellar medium, through the Fischer-Tropsch reaction. UV radiation from the stars subsequently radicalizes the chains, allowing them to turn progressively into aromatic rings, using small, available, hydrocarbon species. Concurrently, the near IR spectrum changes from essentially aliphatic to essentially aromatic.

\end{abstract}


\section{Introduction}
The so-called ``3.4-$\mu$m" band extends from about 3.3 to about 3.6 $\mu$m. It is observed in emission and in absorption towards a wide variety of astronomical objects: DISM (Diffuse InterStellar Medium), reflection nebulae, post-AGB stars, PNe (Planetary Nebulae), PDRs (PhotoDissociated Regions),  HII regions, novae, YSO (Young Stellar Objects). Some of the corresponding spectra are shown in Fig. 1. The shape of the 3.4-$\mu$m band, as well as its strength relative to the 3.3-$\mu$m band are seen to vary considerably. Geballe \cite{geb97} proposed a classification in 4 groups (A to D), taking into account accompanying variations of other UIBs (Unidentified Infrared Bands). Independantly, Tokunaga \cite{tok} defined a very similar classification, differing essentially by the absence of the sparse nova class D. Class A, by far the most populated, has a dominant 3.3 band; several examples can be found in Jourdin de Muizon et al. \cite{dem}. In class B spectra, the 3.4 band intensity is much stronger and may even be dominant; such spectra are observed in post-AGB environments. Although the spectra of novae are weak and short-lived (Hyland and McGregor 1989, Evans et al. 1996), they clearly display both bands with roughly equal strengths. This is also the case for post-AGB stars and PPNe (Proto-Planetary Nebulae) (e.g. here, Fig. 1c; Goto et al. 2003, 2007). These spectra are probably representative of the most pristine carriers of both bands.

In class C, two sharp peaks, near 3.42 and 3.53 $\mu$m, distinctly emerge from a weak plateau covering the same range as the 3.4-$\mu$m band. Such spectra are emitted by YSOs and post-AGB stars.

Geballe \cite{geb97} noted that dominant 3.3-$\mu$m bands are accompanied by conspicuous UIBs at 6.2, 7.7 and 11.3 $\mu$m, while class B spectra have broad emission features, not necessarily reminiscent of the UIBs (see Buss et al. 1993).

More recently, both bands were surveyed across the whole galaxy M82 (Yamagishi 2012). In this survey the whole range of relative intensities, $R=[3.4]/[3.3]$, is covered, from $\ll$1 to $\gg$1. However, the details of the correlative shape variations are hardly distiguishable. Other observations reveal structures in the 3.4 band, which do not appear to be conserved from object to object (see Jourdin de Muizon et al. 1990, Geballe et al. 1985). Sloan et al. \cite{slo} thus distinguished 4 sub-bands, at 3.4, 3.46, 3.51 and 3.56 $\mu$m, in the spectrum of the Orion Bar.

Considering the large number of available near-IR spectra, a general trend seems to emerge: as $R$ increases, one or two narrow sub-bands first arise with a high contrast (see Geballe 1997, Fig. 3); subsequently, the number of sub-bands increases to 4, as their contrast decreases. Ultimately, they merge into a single, wider, band with only a weak structure, as in Fig. 1a (Galactic Center) of the present text or toward the PPN CRL 618 (Lequeux J. and Jourdin de Muizon M. 1990, Chiar et al. 1998) and Seyfert galaxies (see Dartois et al. 2004, Mason et al. 2004). This ultimate shape is generally seen in absorption, and does not appear to vary much between similar objects.

The 3.4-$\mu$m band was quickly interpreted, for it is easily observed in laboratory spectroscopy of organic molecules (hydrocarbons; see Colthup 1990) and solids: hydrogenated amorphous carbon (HAC; see Dischler et al. 1983), kerogens (see Durand 1980), petroleum asphaltenes (Yen et al. 1984), coals (see Painter 1981), and meteorites (see Ehrenfreund et al. 1991). It was assigned to aliphatic CH stretching vibrations of methine (CH), near 3.46 $\mu$m,  asymmetric and symmetric methylene (CH$_{2}$) near 3.42 and 3.48 $\mu$m respectively, and asymmetric and symmetric methyl (CH$_{3}$) near 3.28 and 3.51 $\mu$m respectively (see Yen et al. 1984, Colthup et al. 1990). The extension of this assignment to the astronomical 3.4 $\mu$m band does not seem to be a matter of debate (see Grasdalen and Joyce 1976, Duley and Williams 1981, Papoular et al. 1989, Sandford et al. 1991, Pendleton et al. 1994, Chiar et al. 1998).

The purpose of the present work is to discuss the following issues:

a) What particular chemical structures contribute to the aliphatic feature, possibly also to other UIBs (Unidentified Infrared Bands), but without carrying bands that are not observed in the sky? Can they be distinguished by the frequencies and intensities of their features?

b) Is it possible to simulate the aliphatic band by combining several carriers?

c) How can one explain the very large variations of the intensity ratio $R$, possibly the largest range of relative intensity of any UIB? (see Fig. 1 for objects in our Galaxy; also across galaxy M82, Yamagishi et al. 2012)

d) Can this be related to a coherent trend of dust evolution?

With this in mind, I use computational modeling in Sec. 2, to determine the absorption and emission spectra of several candidate molecules which are also likely to exist in the ISM (InterStellar Medium). Alkane chains, or polymethylenes, (-CH$_{2}$-)$_{n}$ emerge as the preferred choice. In Sec. 3, these spectra are combined with those of more aromatic structures to deliver simultaneously the whole spectrum of CH stretching vibrations, with adjustable relative intensities. Based on these results and on the work of previous authors, Sec. 4 outlines a dust formation scenario which helps understanding some of the trends of $R$ variations observed in space.

\begin{figure}
\resizebox{\hsize}{!}{\includegraphics{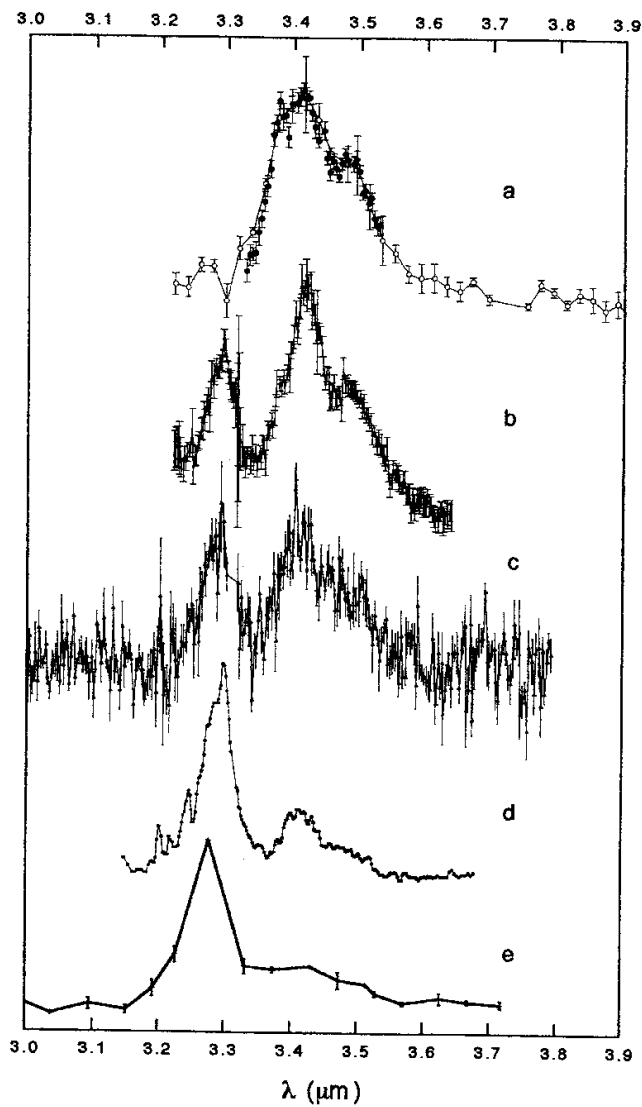}}
\caption[]{The aliphatic C-H stretching bands of a) GC/IRS6E, in absorption (Pendleton et al. 1994); b), c), d)
3 post-AGB stars in emission: IRAS 08341+0852 (Joblin et al. 1996), IRAS 04296+3429 (Geballe et al.
1992) and CRL2688 (Geballe et al. 1992); e) reflexion nebula NGC 2023 in emission (Sellgren 1984).} 
\end{figure}

\section{Radiation absorption and emission by selected candidate carriers}
\subsection{Computational procedure}
The present work is based on the use of various algorithms of computational organic chemistry, as embodied in the Hyperchem software provided by Hypercube, Inc., and described in detail in their publication HC50-00-03-00, and cursorily, for astrophysical purposes, in Papoular \cite{pap01}. Here I use the improved version Hyper 7.5. The computational method used for all molecules is the semi-empirical PM3/RHF algorithm, which is specially tailored for hydrocarbons.

The procedure for any given structure is as follows. First, it is drawn on the computer screen. It is then ``optimized" by minimizing its potential energy at 0 K, a job automatically performed by a sub-routine included in the package. Another sub-routine then performs a Normal Mode Analysis (NMA) to deliver the frequencies and integrated absorption intensities (km.mol$^{-1}$) of all natural vibrations of the given structure. 

The computation of the emission spectrum is more complicated (see Papoular 2012). It is first necessary to choose an excitation mechanism. Here, for purposes of comparison of emission efficiencies, thermal excitation is adopted for all structures: the molecule is embedded in a thermodynamic bath at 300 K for a few picoseconds, enough to give a chance to all vibrations to be excited. The molecule is then allowed to oscillate freely in vacuum at 0 K, for at least 10 ps, while all atomic motions are recorded (so-called Molecular Dynamics run). From these, the total kinetic energy is deduced as a function of time. The Fourier Transform of the latter then delivers the energy distribution among all natural vibrations of the molecule, $E(\nu)$.

Another, identical, molecular dynamics run is then launched, again in vacuum at 0 K, during which the overall dipole moment is monitored. The Fourier Transform of the latter in turn gives the spectral distribution of dipole moment variations, $m(\nu)$. Papoular \cite{pap12} showed that the emission spectrum is given by $W(\nu)=E(\nu)m(\nu)^{2}\nu^{3}$, to a constant factor and in units of energy per frequency interval. This expression is proportional to the excitation energy times Einstein's spontaneous emission coefficient A.

\subsection{Candidate carrier selection}

\begin{figure}
\resizebox{\hsize}{!}{\includegraphics{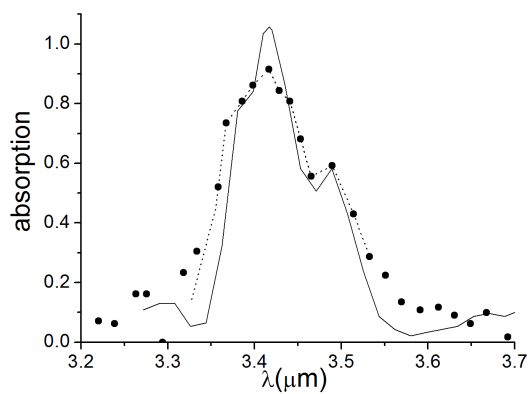}}
\caption[]{Dots and dotted line: extinction spectra of GC/IRS6E, at low and high resolution, respectively (adapted from Pendleton et al. 1994), superimposed upon that of a sample of a lesser evolved coal from the Vouters mine (France), full line; adapted from Guillois 1996.} 
\end{figure}

Non-aromatic structures come with a great variety. Again, the extensive studies of natural carbonaceous substances displaying the spectral features of interest, such as kerogens or unevolved coals, can be of great help in sorting out the most representative members of the family. This is borne out by the excellent fit, in Fig. 2, of the absorption spectrum of a coal sample from the Vouters mine (France), to the extinction spectrum toward the Galactic Center in Fig. 1a. This coal was extracted from a shallow mine, and is therefore akin to unevolved kerogens (cf. Ehrenfreund et al. 1991, for fits to other natural materials). Its composition is characterized by the following atomic ratios: H/C$\sim0.7$, O/C$\sim0.09$, and its structure includes benzenic rings, single or in clusters.

The presence of kerogen-like material in carbonaceous chondrites strongly suggests that kerogen might be a good model for interstellar dust. It has been shown (see Behar and Vandenbroucke 1986; Vandenbroucke 1980), by chemical analysis and simulation that the lesser evolved (aromatized) kerogens are essentialy made of straight or branched chains, with oxygen bridges, and naphtenes (chains of cycloalkanes, such as cyclohexanes, C$_{6}$H$_{12}$). As aromatization proceeds with age and depth underground, aromatic cycles form and coalesce between chains, in compact (pericondensed) clusters generally not exceeding a handful of rings. We are not interested, here, in the later stages of aromatization, when the benzenic clusters become dominant at the expense of the aliphatic structures.

We recall that previous authors, based on other grounds, also invoked these types of structures (see Schutte et al. 1993, Bernstein et al. 1996, Joblin et al. 1996, Chiar et al. 1996).

Note that aliphatics also include alkenes and alkynes (doubly and triply bonded carbon chains; polyynes and cumulenes). However, these structures do not occur frequently in kerogen models; moreover, they exhibit vibrational features near 3 and 4 $\mu$ which are not conspicuous in astronomical spectra, so they will not be considered further here.  

The structures that were selected for computation in the present study are represented in Fig. 3 to 10. 

\begin{figure}
\resizebox{\hsize}{!}{\includegraphics{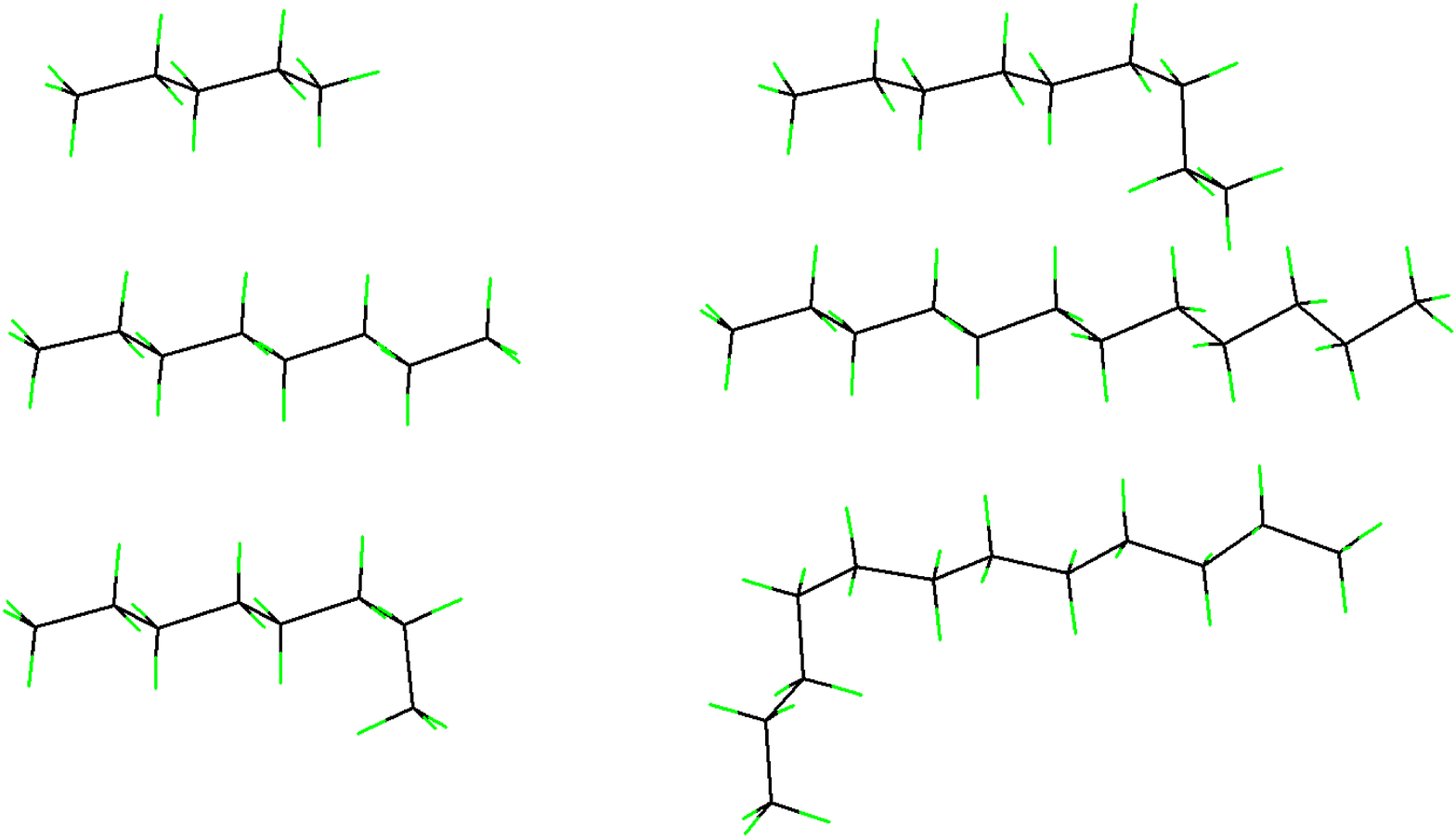}}
\caption[]{Short, straight or bent, alkane chains; 5, 8, 8(bent), 9(bent), 12, 12(bent) C atoms. Black: carbon; green: hydrogen.
Color on-line.} 
\end{figure}

\begin{figure}
\resizebox{\hsize}{!}{\includegraphics{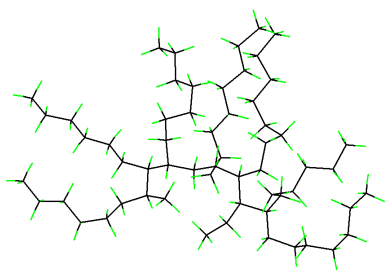}}
\caption[]{Branched aliphatic (alkane) chains. Black: carbon; green: hydrogen. 59 C, 120 H atoms. Color on-line.} 
\end{figure}
\begin{figure}
\resizebox{\hsize}{!}{\includegraphics{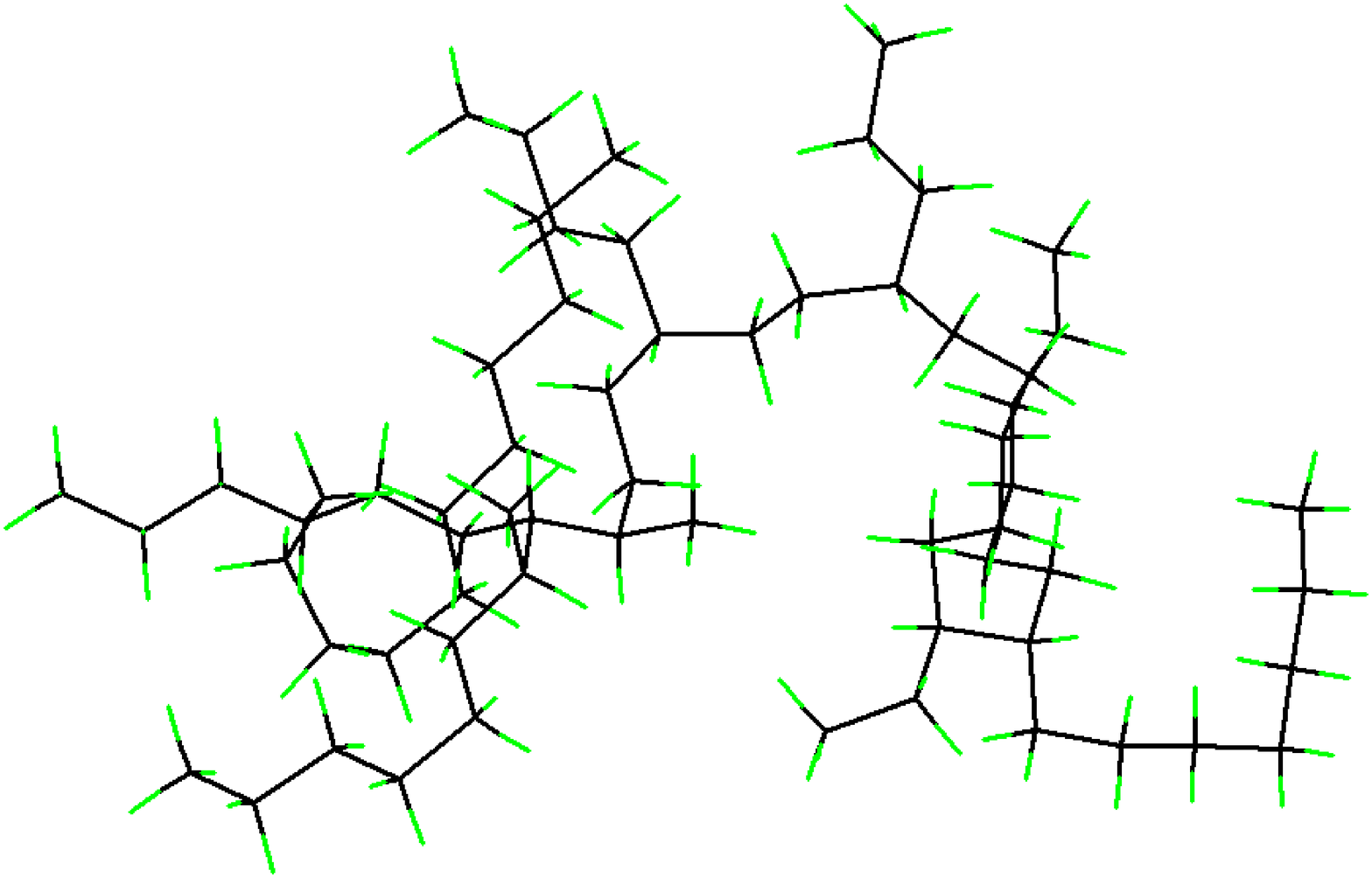}}
\caption[]{Another version of aliphatic chain structure. 63 C, 128 H atoms. Color on-line. } 
\end{figure}

\begin{figure}
\resizebox{\hsize}{!}{\includegraphics{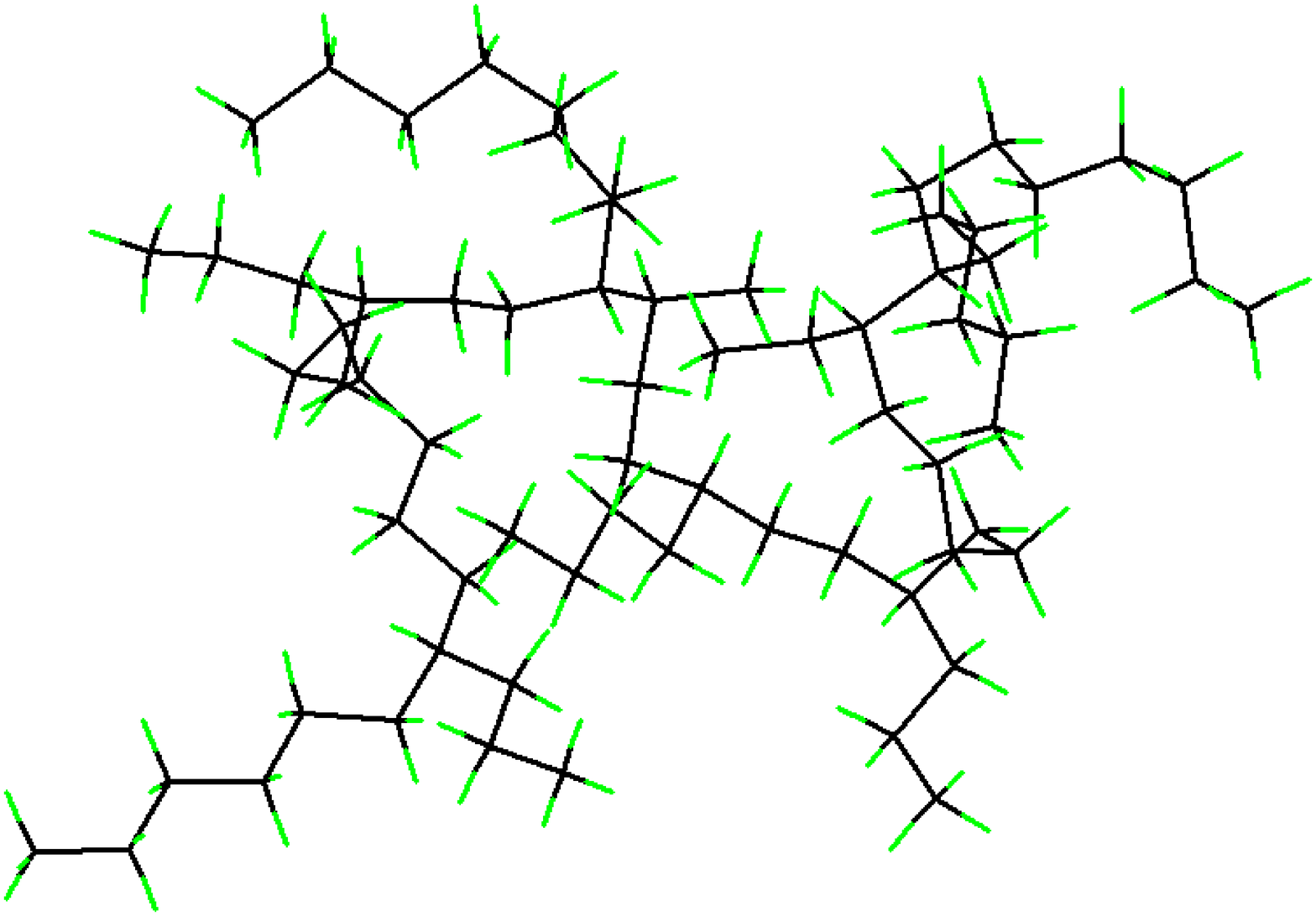}}
\caption[]{A third version of aliphatic chain structure. 69 C, 140 H atoms. Color on-line. } 
\end{figure}

\begin{figure}
\resizebox{\hsize}{!}{\includegraphics{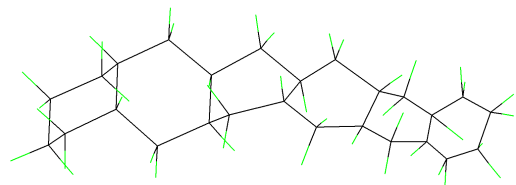}}
\caption[]{Naphtenic structure composed of catacondensed cyclohexane rings. 26 C, 42 H atoms. Color on-line. } 
\end{figure}
\begin{figure}
\resizebox{\hsize}{!}{\includegraphics{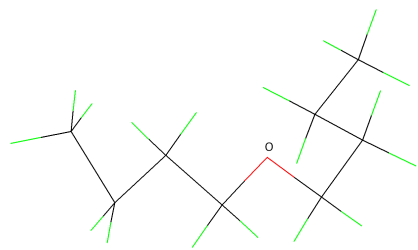}}
\caption[]{Alkane chains connected by an oxygen bridge (red). 8 C, 18 H, 1 O atoms. Color on-line. } 
\end{figure}
\begin{figure}
\resizebox{\hsize}{!}{\includegraphics{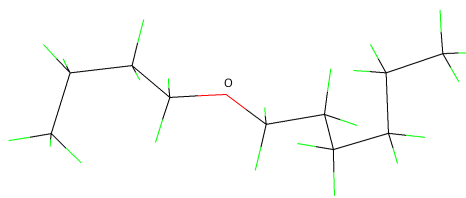}}
\caption[]{Another version of O-bridged alkane chains. 10 C, 22 H, 1 O atoms. Color on-line. } 
\end{figure}
\begin{figure}
\resizebox{\hsize}{!}{\includegraphics{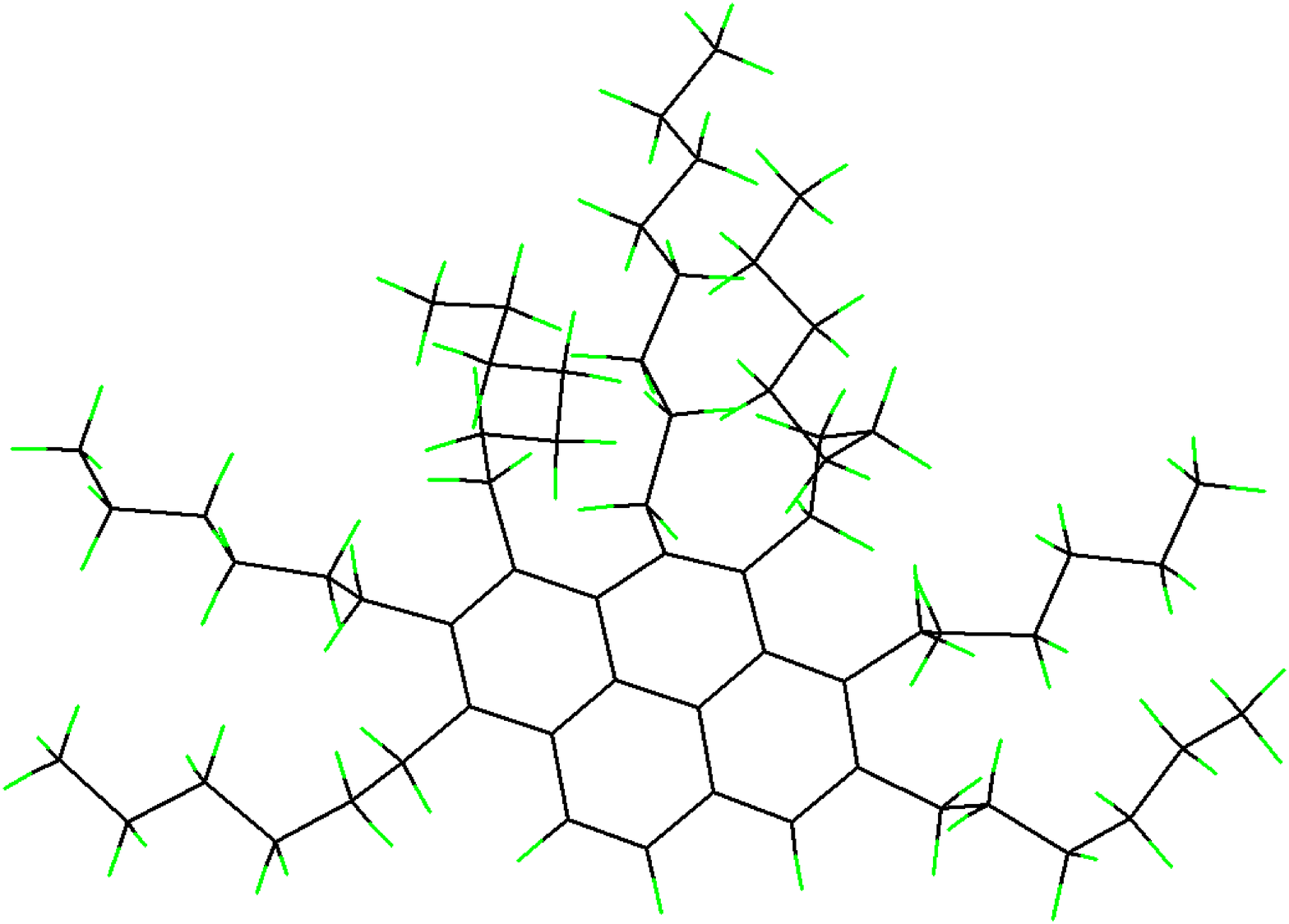}}
\caption[]{Combined aliphatic (alkane chains) and aromatic (pyrene) structure. 63 C, 104 H atoms. Note the 3 single CH bonds, attached to the pyrene molecule, which provide the partially aromatic character. Color on-line.  } 
\end{figure}

\subsection{IR absorption spectra}

Near IR integrated absorption spectra of short alkane chains, as delivered by NMA, are displayed in Fig. 11. The lines bunch in 4 groups, near 3.15, 3.25, 3.3 and 3.4 $\mu$m. The simulation package allows one to animate individually each vibration mode of any given structure, so their nature can be determined. It is thus found that the 4 groups correspond, respectively, to symmetric, asymmetric methyl, symmetric and antisymmetric methylene stretching vibrations. The CH$_{3}$ vibrations are intrinsically weak and their occurrence is rare in kerogens and meteorites, as they serve essentially as terminations for alkane chains, so their contribution is negligible. Even the CH$_{2}$ vibrations are not strong. However, it appears that bending a chain tends to strengthen its IR activity.

Figure 12 displays the superposed spectra of the three branched-chains structures, together with that of the 9-C atoms chain of Fig. 3, (x10), for comparison. Obviously, branching considerably strengthens the line intensities. Moreover, the attending increase in number of atoms and vibration modes densely populates the aliphatic stretching spectral window, filling the intervals between the 4 groups of Fig. 11. Associating many small modifications of these 3 structures will ultimately give rise to a continuous band. It is therefore reasonable to surmise that a combination of short and branched CH$_{2}$ chains can yield a continuous band more or less modulated by the distinct subbands associated with short chains (see emission spectra, below).

In assessing these spectra, it must be kept in mind that the vibrational frequencies delivered by chemical simulation codes are generally too high by 1 to 10 $\%$, depending on the type of vibration (see Hehre et al. 1993). I have compared frequencies computed using the Hyperchem package with the measured frequencies of several molecules. The maximum error of the code used in this work occurs for the symmetric methyl vibration (all 3 CH bonds in phase): $\sim10\%$. Fortunately, methyls contribute very little to our spectra. For the asymmetric methyl vibrations, the error is of order $1\%$ ($\Delta\lambda\sim0.03 \mu$m), as is the case for the other structures. When these defects are taken into account, Fig. 11 and 12 may recall the main characters of the type A and B astronomical aliphatic bands, between 3.35 and 3.6 $\mu$m, with a prominent peak near 3.4 $\mu$m.

\begin{figure}
\resizebox{\hsize}{!}{\includegraphics{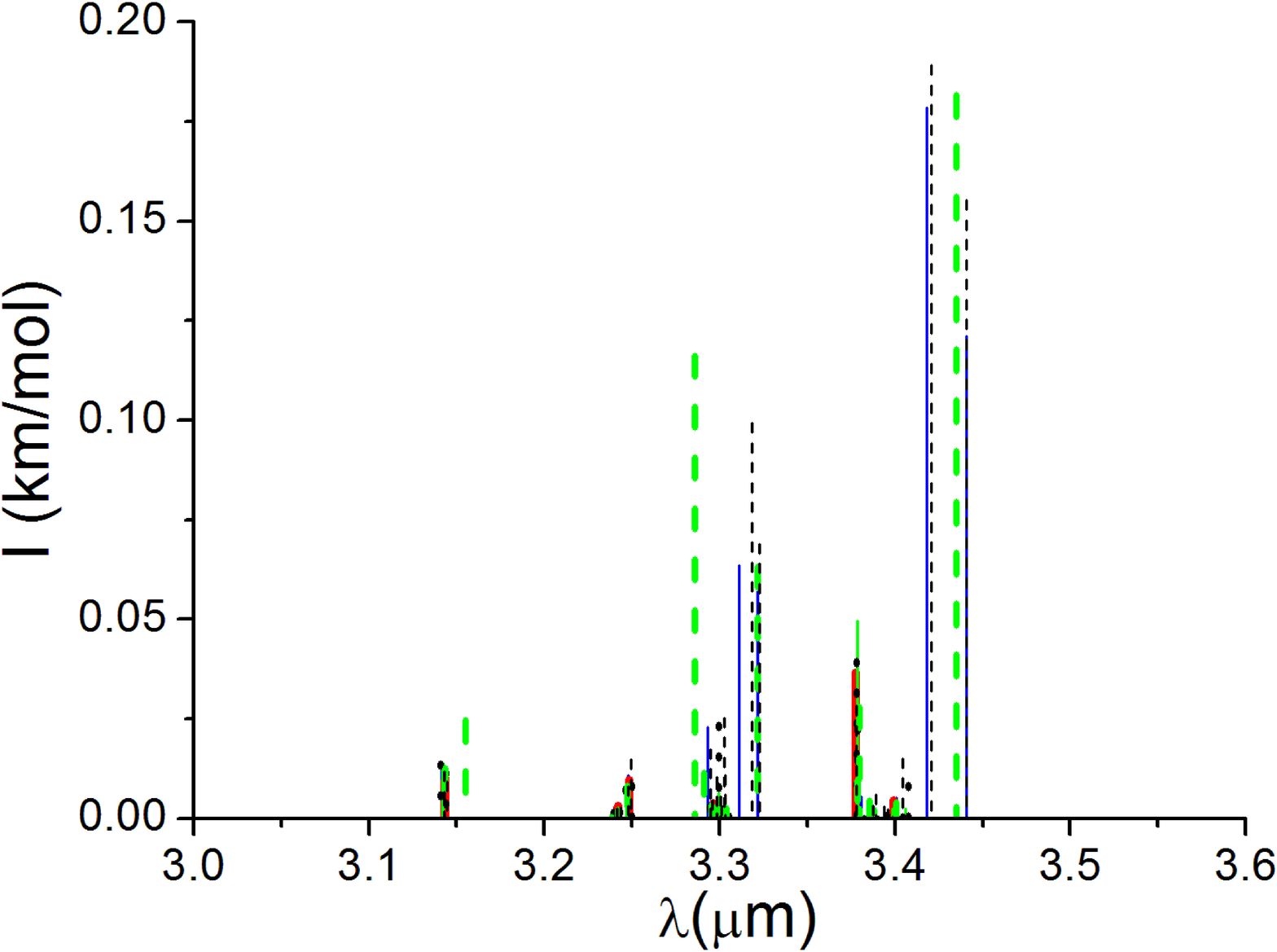}}
\caption[]{The near IR absorption spectra (integrated intensities or band areas) of the (short) alkane chains of Fig. 3, in the same order; respectively: red (very weak), green, dashed green, blue, black, dotted black. Color on-line.} 
\end{figure}

\begin{figure}
\resizebox{\hsize}{!}{\includegraphics{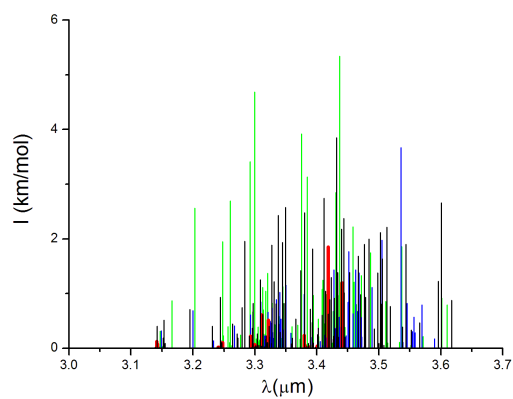}}
\caption[]{The near IR absorption spectra (integrated intensities) of the branched alkane chains of Fig. 4 (green), 5 (blue) and 6 (black) together with that of the short, straight alkane chain at the top left of Fig. 3, for comparison (red; from Fig. 11). The intensities of the latter were multiplied by 10 for clarity: it appears that the spectral intensities of branched chains are much stronger. Color on-line. } 
\end{figure}

\begin{figure}
\resizebox{\hsize}{!}{\includegraphics{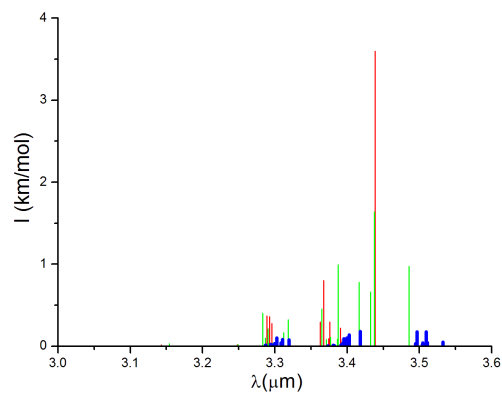}}
\caption[]{The near IR absorption spectra (integrated intensities) of the other structures, Fig. 7 to 9, respectively in blue, red and green. These structures are less IR-active than branched chains (Fig. 12), except near 3.4 $\mu$m, where Oxygen is seen to considerably increase the strength of attached short, straight CH$_{2}$ chains. Color on-line. } 
\end{figure}

\begin{figure}
\resizebox{\hsize}{!}{\includegraphics{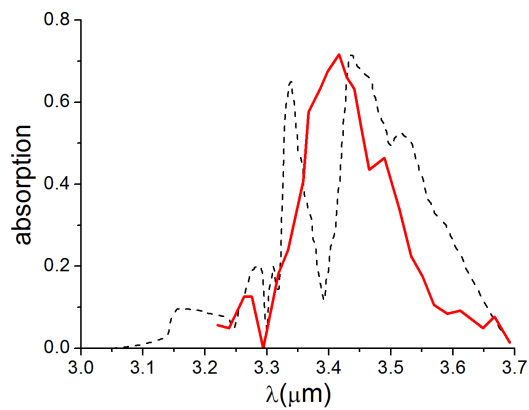}}
\caption[]{Red line: extinction spectrum of GC/IRS6E, at low resolution (adapted from Pendleton et al. 1994). Black dashes: model absorption spectrum synthesized by concatenating the spectra of the following structures: short chain 9(bent) (Fig. 3), branched chains of Fig. 5 and 6 and naphtene (Fig. 7). This was then smoothed by FFT filtering over 0.1 $\mu$m. Color on-line. }  
\end{figure}

The main aliphatic vibrations of the other structures fall in the same spectral window (see Fig. 13). These structures are less IR-active than branched chains (Fig. 12), except near 3.4 $\mu$m, where Oxygen is seen to considerably increase the IR strength of attached short, straight CH$_{2}$ chains.

\begin{figure}
\resizebox{\hsize}{!}{\includegraphics{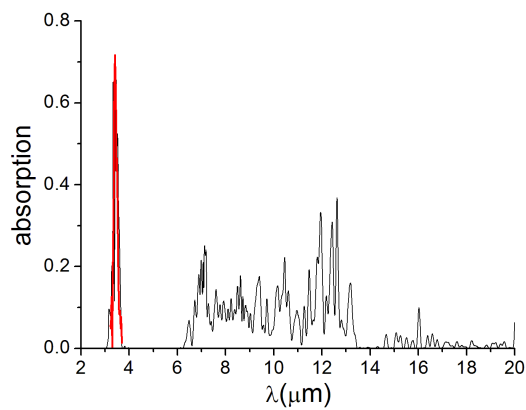}}
\caption[]{Extension of the model spectrum of Fig. 14, to show the weakness of the mid IR features relative to the near IR stretching band.}
\end{figure}

As an example of the synthetic absorption spectra that can be obtained using these elementary spectra, the dashed line in Fig. 14 represents the near IR absorption spectrum obtained  by combining the spectra of the bent chain (9bent) in Fig. 3, and the branched chains of Fig. 5 and 6, with the spectrum of naphtene (Fig. 7). The IRS6E low resolution spectrum of Fig. 2 is superimposed for comparison. The spectrum of the branched chains in Fig. 4 was excluded from this synthetic spectrum because it carries too much IR activity below 3.35 $\mu$m (see Fig. 12), as were the spectra of O-bridged chains because they carried too much activity near 6 and 13 $\mu$m. 

In assessing this tentative ``fit", it must be observed that shifts of 1 or 2$\%$, well within avowed simulation errors, would suffice to bring the main peaks of the model spectrum in coincidence with those of IRS6E. Such shifts could also be induced by anharmonicity effects which are not accounted for by NMA in absorption spectra (Papoular 2012). The computed emission spectra shown in the next Section (Fig. 18 to 20) are much better models of astronomical spectra.

Finally, the handful of component spectra retained for this synthesis certainly do not exhaust the list of all the astronomical carriers, but this fitting exercise clearly shows what structural types are favoured, so that it is straightforward to go further by increasing the number of model structures and refining the tailoring of synthetic spectra. However, again, one is restricted in this intent by the insufficient accuracy of the computed frequencies and intensities. 

Figure 15 displays the mid IR part of the same spectrum as Fig. 14. It is apparent that the selected structures contribute IR features in the same spectral window as the Murchison meteorite (see de Vries et al. 1993) albeit with much less intensity relative to the CH stretching band. This is in line with the weakness of astronomical mid IR extinction accompanying the 3.4 $\mu$m band (see Mason et al. 2004). Still, the features near 7 $\mu$m detected by Chiar et al. \cite{chi00} may have something to do with those of the present model, in Fig. 15.

Note that the absence of short, straight chains or O-bridged chains in the list of component spectra for this model does not mean that they are not present in the sky: the model only suggests that, relative to branched chains, their number in space is not large enough to compensate for their weak near IR activity.

\subsection{IR emission spectra}

Figure 16 and 17 display the computed near and mid IR emission spectra of the selected individual structures: a) short chain 9bent (Fig. 3), b) c) d) branched chains (Fig. 4 to 6), e) naphtene (Fig. 7), f) and g) O-bridged chains (Fig. 8 and 9). As in absorption, the branched chains dominate over all other spectra. The same comment about the mid IR spectra, and the wavelength accuracy, can be made as for absorption spectra.

Figure 18 is an attempt at simulating an extreme type B (purely aliphatic) emission spectrum from Fig. 2 of Yamagishi et al. \cite{yam} or that in absorption of NGC 1068 from Mason et al. \cite{mas}. Here the combination of spectra (a) to (g) is:
1x(a)+3x(b)+3x(c)+1x(d)+10x(e)+1x(f)+1x(g). The profile of this emission spectrum is a much better model for type B than is Fig. 14.

\begin{figure}
\resizebox{\hsize}{!}{\includegraphics{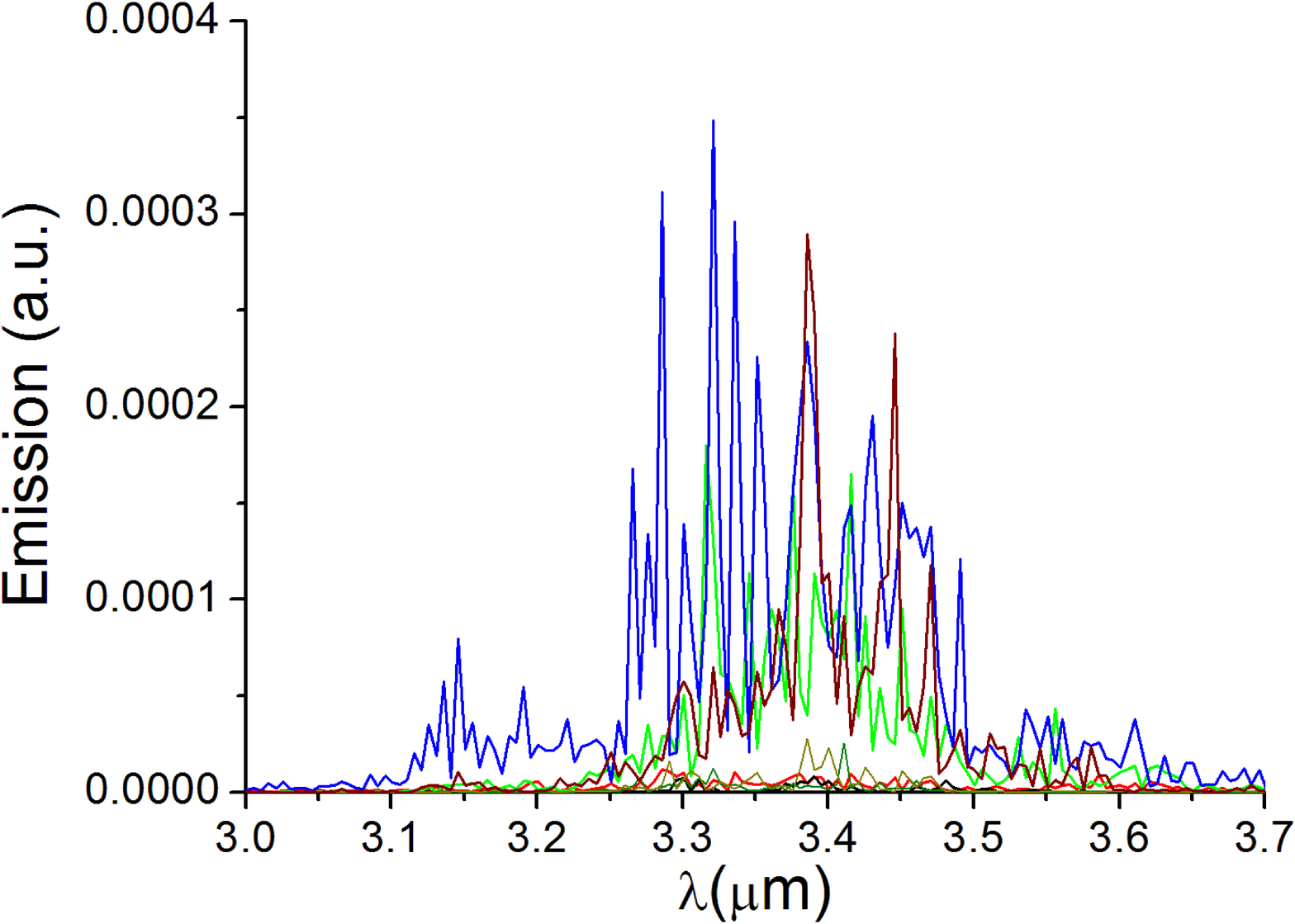}}
\caption[]{Near IR emissions at 300 K. Red: short chain 9bent, Fig. 3. Green, blue and wine: structures in Fig. 4 to 6. Black: Fig. 7. Olive: Fig. 8. Dark yellow: Fig. 9. Color on-line.} 
\end{figure}

\begin{figure}
\resizebox{\hsize}{!}{\includegraphics{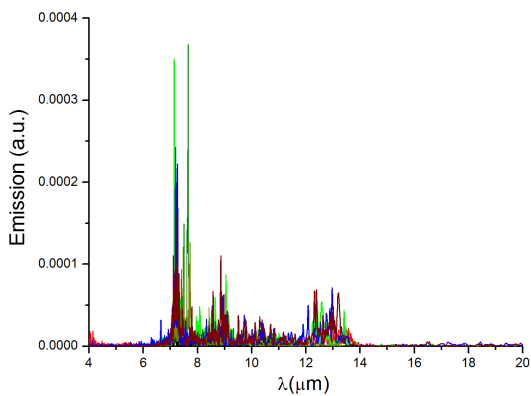}}
\caption[]{Mid IR emissions at 300 K. Red: short chain 9bent, Fig. 3. Green, blue and wine: Fig. 4 to 6. Black: Fig. 7. Olive: Fig. 8. Dark yellow: Fig. 9. Color on-line.}  
\end{figure}

\begin{figure}
\resizebox{\hsize}{!}{\includegraphics{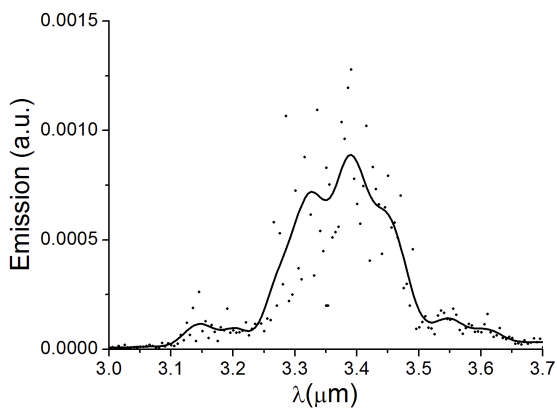}}
\caption[]{A tentative simulation of an extreme class B emission spectrum at 300 K, in the near IR (see Sec. 2.4). Dots: computed points. Line: after smoothing over 0.025 $\mu$m. Note the similarity with emission spectra from the halo of M82 (Fig.2. of Yamagishi et al. 2012), and with the absorption spectrum of the Galactic Center in Fig. 2 above.} 
\end{figure}

\section{Combined aliphatics and aromatics}
Models of kerogens indicate that aliphatic chains are often attached to small clusters of pericondensed aromatic rings, as in Fig. 10. Figure 19 represents the emission spectrum of this structure at 300 K. It highlights the strong IR activity of aromatic CH stretching: only 3 aromatic CH bonds are enough to produce a much stronger band than the 101 aliphatic CH bonds.

\begin{figure}
\resizebox{\hsize}{!}{\includegraphics{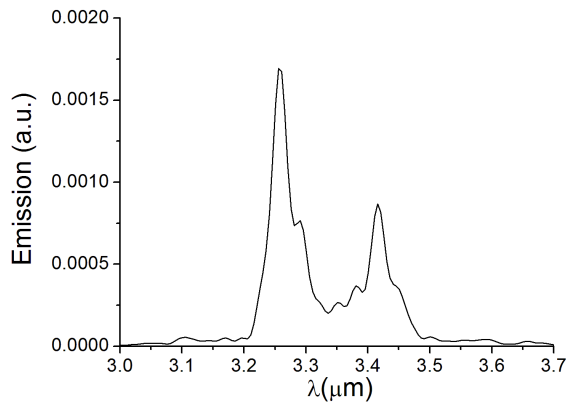}}
\caption[]{The computed emission spectrum of the  composite aromatic/aliphatic structure of Fig. 10, in the near IR. Smoothed over 0.013 $\mu$m. The aromatic feature emitted by only 3 aromatic CH bonds is much stronger than the aliphatic feature delivered by the 101 aliphatic CH bonds of the same molecule.}  
\end{figure}

\begin{figure}
\resizebox{\hsize}{!}{\includegraphics{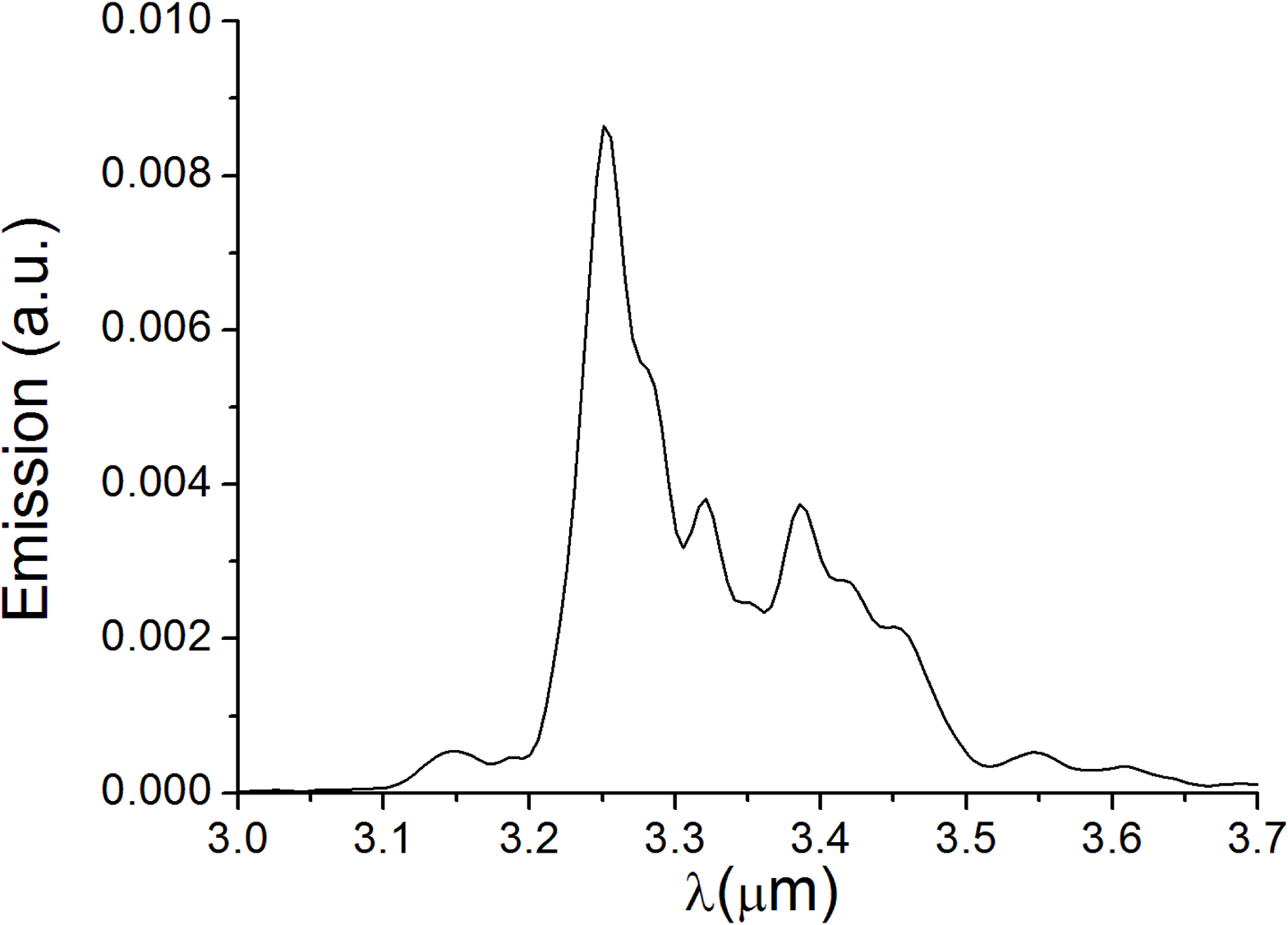}}
\caption[]{The combination of a mainly aromatic (UIB-like) and a mainly aliphatic synthetic emission spectra, both computed (see Sec. 3). Smoothed over 0.013 $\mu$m. Note the similarity with emission spectra from the disk of M82 (Fig.2. of Yamagishi et al. 2012).} 
\end{figure}

Another example of mixed types is provided by Fig. 20. Here, the aliphatic spectrum of Fig. 18 was strengthened by a factor 10, and added to the mainly aromatic synthetic emission spectrum computed along the same lines in Papoular 2012 (Fig. 9), although the purpose, there, was to simulate a typical UIB spectrum. The result of this combination is comparable to the spectra from the disk of M82, gathered by Yamagishi et al. \cite{yam}, given the inaccuracy of the simulation code. By adjusting the multiplying factor, one can cover the whole observed range of aliphatic to aromatic ratio, $R$, delivered by M82.

\section{Formation and evolution scenario}

If one is willing to draw, from the above, the conclusion that the main carriers of the aliphatic stretching band are short and branched CH$_{2}$ chains, then, the ubiquity, and sometimes the prevalence of this band over the aromatic stretching band, underscore the importance of these structures as  components of interstellar dust. Also, there seems to be a continuous thread between aliphatic and aromatic structures, which parallels the evolution of astronomical near IR spectra from high to low $R$. There are reasons to believe that aliphatic chains are the precursors of more elaborate carbonaceous dust, aromatics in particular. For it is very difficult to imagine, instead, a process by which the strongly bound aromatic rings can be continuously split up into chains. Besides, ``PAH spectra" have been observed to emanate from very harsh environments (see Yamagishi et al. 2012). By contrast, processes are known in the laboratory and in industry, which can synthesize chains from simple molecules common in the ISM, and then build aromatic rings from such chains, as shown below.

The Fischer-Tropsch (FT) reaction,\\

(2n+1)H$_{2}$+nCO $\rightarrow$ C$_{\mathrm{n}}$H$_{2\mathrm{n}+2}$+n H$_{2}$O\,,\\

also described as ``CO polymerization'', is known and used industrially to produce preferentially straight alkane chains, the progenitors of the most efficient 3.4-$\mu$m carrier, as shown in Sec.~2.

While, in industry, several preliminary steps are required before this reaction may take place, the atmospheric conditions in CircumStellar Envelopes  (CSE) are obviously naturally favorable, H$_{2}$ and CO being the most abundant molecules in such envelopes. The average temperature is also close to that under which the industrial process is conducted (150 to 300 °C). Even the most common catalyst used in industry, iron, is also available in CSE.

Of course, stars with strong winds are the more propitious to such a reaction, and this is the case for AGB stars and novae, in particular. This is illustrated by Fig. 1b and 1c above, and by Nova Cen 1986 (Hyland and McGregor 1988). Several authors were thus led to the conclusion that the aliphatic carriers are born in such environments (see Lequeux and J. de Muizon 1990, Goto et al. 1997).

Now, how can aliphatics convert to aromatics along the trend suggested by Fig. 1 ? The study of flames in the laboratory is very instructive in this respect (see Harris and Weiner 1985, among others). Building upon the extensive work of Wang and Frenklach \cite{wan} on PAH growth in flames, Frenklach and Feigelson \cite{fre} aptly suggested that this conversion may be prompted by radicalization (under heating or irradiation), as in the following succession of steps:

\begin{flushleft}
C$_{edge}$H+H $\rightleftharpoons$ C$_{edge}$*+H$_{2}$\\
C$_{edge}$*+H $\rightarrow$ C$_{edge}$H\\
C$_{edge}$*+hydrocarbon growth species$\rightarrow$growth of aromatic rings,
\end{flushleft}

where the asterisk designates a radical, and the growth species include C$_{2}$H$_{2}$ (preferred), CH$_{3}$, C$_{2}$H$_{3}$, C$_{3}$H$_{3}$, etc. which are quite common in space (see Goto et al. 2003, Chiar et al. 1998). Many of the latter are expected to form along with the FT reaction. This path consists essentially in H-abstraction followed by addition of a small hydrocarbon molecule. 

According to Frenklach and Feigelson \cite{fre}, in terrestrial combustion and laboratory pyrolysis, the rates of aromatics production peak between 1600 and 2000 K. In CSE, however, atmospheric conditions are not ideal, as the ejected gas expands and cools promptly. This may explain why the 3.3-$\mu$m band is seldom observed at the edge of CSE of RGB or AGB stars. Instead, it becomes prominent when the central star is unveiled and very hot, as in the generic Orion Bar and Red Rectangle, for instance. Rather than high temperatures and shocks, this is due to the strong stellar UV radiation, which dissociates gaseous molecules, thus creating a PDR (PhotoDissociated Region) near by, and providing the required radicals. Obviously, the aromatizing process can also take place in the Diffuse Interstellar Medium and in Reflection Nebulae (cf. Fig. 1e), albeit more slowly, because of the weaker radiation density.

In this context, it may be recalled that, in the laboratory as well, heat is not enough to aromatize carbons. It was thus found that aromatization requires a significant amount of ``roaming" hydrogen atoms (or `` mobile phase": see Given et al. 1986). Otherwise, the carbon material is ``non-aromatizing". 

The increase by a factor 7 of the band ratio R, from the center to the edge of M82 (Yamagishi et al. 2012, their Fig. 3) can be explained along the same lines. Indeed, the 7-$\mu$m band (5.9 to 8.4 $\mu$m) measured by the AKARI satellite (see Yamagishi et al. 2012) decreases in intensity by 3 orders of magnitudes from the center to the edge of the galaxy, suggesting a similar trend for the UV radiation, such as may result from a decreasing spatial density of hot and luminous stars. As a consequence, aromatization is expected to proceed much more slowly, if at all, with increasing distance from the center. As a matter of fact, at the galaxy edge (halo), the 3.3-$\mu$m band is dwarfed by the 3.4-$\mu$m band (Yamagishi et al. 2012, Fig. 2).

Looking at ordered sequences of astronomical spectra, such as in Fig. 1, one notes that, as aromatization progresses, not only does the aliphatic band weakens relative to the aromatic one, but it also becomes more bumpy and sometimes ultimately exhibits a few secondary peaks rising distinctly above a weak plateau (see Sloan et al. 1997 and class C spectra). In the scenario proposed here, this may be explained as follows. Since aromatization proceeds at the expense of aliphatic chains, it is plausible that chains decrease in length as a consequence. Now, Fig. 11, 12 and 16 indicate that, in emission as well as in absorption, the spectra of short chains are rather weak and discrete, with two or three prominent subbands, as opposed to those of long, branched chains, which extend rather uniformly across the aliphatic window. It would seem, then, that the aliphatic bands of class B spectra are carried mostly by long, branched chains; upon aromatization, the latter progressively turn into short, straight chains, which carry the residual aliphatic band in class A, and, ultimately, in class  C spectra.
                                                                                          Finally, one may interpret the predominance of the aliphatic band in absorption towards the center of our own Galaxy (see e.g. Fig. 1, top) as due to the abundance of RGs and AGBs, and dirth of atomic hydrogen, along that line of sight, precluding aromatization but favoring aliphatic chains. The fact that a few sub-peaks still modulate the spectral profile suggests that both short and branched chains contribute to such spectra.

\section{Discussion}
Li and Draine \cite{lia} recently commented about the fraction of aliphatic bonds present in the carriers of UIBs. Based on the averaged observed ratio of the peak intensities of the 3.4- and 3.3-$\mu$m features in UIB spectra, which they take to be 0.2, and the ratio of the corresponding integrated band intensities, which they take to be about 0.7, they conclude that there are no more than 9 $\%$ of aliphatic bonds in the carriers. This argument raises a few  questions.

1) While the ratio of 0.2 for the peak intensities is indeed observed in the most aromatic cases, the very large variations of this ratio according to environment and evolutionary stage is documented in the Introduction above, as is recalled the Geballe and Tokunaga classifications in 4 or 3 types. Examples of these are seen in Fig.1. This variety is conspicuous, for instance, in the spectral maps of M82 \cite{yam}. It appears, therefore, that much larger values have also to be accounted for.

2) The notion of integrated absorption ``intensity per bond" is certainly valid for extremely localized vibrations and very well defined functional groups, and only for transitions between the ground and first excited states. In general, however, chemists define integrated intensities \it per vibrational mode of the whole molecule. \rm This is because, in larger molecules, and in emission, functional groups all over the molecule are coupled together by anharmonicity, which may considerably affect their respective ``efficiencies". One classical example is the influence of even a few peripheral OH groups on the other spectral features of the molecule. Another example is the complex molecule drawn in Fig. 10: it carries only 3 aromatic CH bonds (attached to the pyrene cluster) and 101 various aliphatic  bonds and, even so, the 3.3-$\mu$ feature is much stronger than the 3.4-$\mu$ one, as Fig. 19 shows.

3) In astronomical emission spectra, the width of the 3.4-$\mu$ band is generally much larger than that of the 3.3-$\mu$m band. In that case, the ratio of band areas seems to be a better estimate of the ratio of emitted energies than is that of the peak intensities.

It appears, therefore, that the best way to characterize the carriers of a given astronomical spectrum would be to try and simulate this particular spectrum, as best as possible, with known materials, in the laboratory or by chemical modeling, in emission and/or absorption. This concurs with the conclusion of Li and Drain \cite{lia}.

\section{Conclusion}

The approach of this work to carbonaceous IS dust owes a great deal to the methods and findings of experts in oil, kerogen and coal. From the extensive efforts to model these materials it appears that this is not possible with a finite number of finite molecules. They suggest, instead, an infinite number of medium sized molecules falling in a small number of structural classes, within each of which the component molecules differ only slightly in size and composition. However, acceptable models can be built with a reasonable number of components and even with available computational means, remembering that each model combination of components applies only to a given terrestrial material.

Similarly, in space, depending on location and antecedents, dust models should be tailored with different combinations of components to fit different spectral profiles along different sightlines. In the present work, models of the 3.4-$\mu$m band were obtained with various combinations of 7 mainly aliphatic structures: -CH$_{2}$- straight and bent chains, some including oxygen bridges or small pericondensed PAHs like pyrene, some of them branched together, and, finally, naphtene chains. The whole near-IR stretching band, i.e. the 3.3-$\mu$m band included, can be modeled by adding to these a suitable combination of 21 mainly aromatic structures made of C, H, O, N and S atoms, as described in a previous work dedicated to the simulation of the UIBs. By varying the relative weights of the aliphatic and aromatic components, it is possible to obtain reasonably good fits to generic astronomical spectra displaying different ratios of 3.3/3.4 intensities and different degrees of profile modulation/smoothness.

\end{document}